# Low-temperature infrared spectroscopy of $Tm_{0.19}Yb_{0.81}B_{12}$ dodecaboride with metal-insulator transition and dynamic charge stripes.

E.S.Zhukova[1,2], A.Melentyev[1], B.P.Gorshunov[1,2], A.V.Muratov[3], Yu.A.Aleshchenko[3], A.N.Azarevich[2,1], K.M.Krasikov[1,2], N.Shitsevalova[4], V.Filipov[4], N.E.Sluchanko[2,1]

[1]Moscow Institute of Physics and Technology, 141700 Dolgoprudny, Moscow Region, Russia

[2]Prokhorov General Physics Institute of the Russian Academy of Sciences, Vavilov str. 38, Moscow 119991, Russia

[3]Lebedev Physical Institute, Russian Academy of Sciences, 119991 Moscow, Russia

[4]Frantsevich Institute for Problems of Materials Science, National Academy of Sciences of Ukraine, 03680 Kiev, Ukraine

**Abstract.**

$Tm_{1-x}Yb_xB_{12}$ dodecaborides represent model objects for the studies of quantum critical behavior, metal-insulator transition and complex charge-spin-orbital-phonon coupling phenomena. In spite of intensive investigations, the mechanism of the ground state formation in this strongly correlated electron system remains a subject of active debates. We have performed first systematic measurements of temperature-dependent spectra of infrared conductivity of $Tm_{0.19}Yb_{81}B_{12}$ at frequencies 40-35000 cm$^{-1}$ and in the temperature interval from 300 K down to 10 K. Analysis of the temperature evolution of the observed absorption resonances is performed. Their origin is associated with the Jahn-Teller instability in the cubic lattice which results in the rattling modes of the rare earth ions and leads to emergence of both the intra-gap mixed-type collective excitations and the dynamic charge stripes. Temperature-dependent effective mass of charge carriers is determined and a picture is presented of the multiple relaxation channels and the transformation of the manybody states at different temperatures. We argue in favor of electronic phase separation scenario which is valid both for the metal-insulator transition and for the formation of the nanoscale filamentary structure in $Tm_{1-x}Yb_xB_{12}$ compounds.

**Introduction.**

Among rare-earth dodecaborides, the solid solutions $Tm_{1-x}Yb_xB_{12}$ are of special fundamental interest due to a metal-insulator transition (MIT) that takes place in these strongly correlated electron systems when thulium is substituted by ytterbium. With the increase of ytterbium content from $x=0$ ($TmB_{12}$, $4f^{12}$ configuration) to $x=1$ ($YbB_{12}$, $4f^{13+\delta}$ configuration, $\delta=0$-0.05 [1]), the properties of the $Tm_{1-x}Yb_xB_{12}$ gradually transform from the antiferromagnetic in metallic $TmB_{12}$ ($T_N=3.2$ K [2]) to the paramagnetic in the narrow-gap Kondo insulator $YbB_{12}$ with highly correlated charge carriers [3-5]. The MIT leads to a considerable growth at helium temperatures of the DC resistivity, from 4 µΩ cm in $TmB_{12}$ to 10 Ω cm in $YbB_{12}$, changing its behavior from metal-like to semiconductor-like [6,7]. In spite of intensive investigations, the nature of this nonmagnetic semiconducting state remains a subject of debate [8-10]. There are several models proposed to explain the insulating nature of $YbB_{12}$. One of them is based on the coherent band picture where the energy gap is formed due to the strong c-f hybridization [11,12]. The inversion between 4f and 5d bands that accompanies this process is considered as an essential characteristic of the topological Kondo insulator [13-15]. Other models are based on a local picture, where the electrons contributing to the Kondo screening are captured by the local magnetic moments resulting in an excitonic local Kondo bound state [16,17]. The single-site scenarios were proposed also by Liu [18] and by Barabanov and Maksimov [10].

Recent Hall effect experiments on $Tm_{1-x}Yb_xB_{12}$ reveal complex activated behavior of charge carrier concentration and non-monotonous temperature dependence of their mobility [19]. Contrary to all the above mentioned models, the charge carrier transport was interpreted in [6] in terms of the electron phase separation effects in combination with the formation of nanosize clusters of rare earth (RE) ions and emergence of Yb–Yb pairs in the $Tm_{1-x}Yb_xB_{12}$ cage-glass matrix [19]. In addition, with the increase of $x$(Yb), antiferromagnetic instability was found to develop in the $Tm_{1-x}Yb_xB_{12}$ compounds with the quantum critical point near the critical concentration $x_c \approx 0.25$ [6,7]. Recently, based on accurate low temperature transport measurements combined with high-precision x-ray diffraction experiments on single-domain crystals, formation of highly conducting dynamic charge stripes within the semiconducting matrix of $Tm_{0.19}Yb_{0.81}B_{12}$ dodecaboride was discovered and characterized [20]. Room-temperature infrared spectroscopic measurements allowed the authors to estimate the frequency (~$2.4 \times 10^{11}$ Hz) of quantum motion of the charge carriers within these stripes. Two peaks observed at 107 cm$^{-1}$ and 132 cm$^{-1}$ were associated with the large-amplitude quasilocal vibrations of the Yb and Tm ions which are caused by the cooperative dynamic Jahn–Teller effect in the boron sub-lattice. These are known as rattling modes (Einstein oscillators) responsible for variation of 5$d$-2$p$ hybridization of electron states leading to the 'modulation' of the conduction band and formation of dynamic charge stripes along one of the <110> directions (see insert in Fig.1a). A number of other excitations were observed in the range $10 - 10^4$ cm$^{-1}$ whose origin was not identified. It was found that intrinsic "inhomogeneity" of the crystal due to the charge stripes led to an extremely narrow Drude peak in the spectra and correspondingly very small, "artificial" scattering rate of the charge carriers $\gamma_D \approx 8$ cm$^{-1}$ [20].

In this regard, $Tm_{1-x}Yb_xB_{12}$ dodecaborides with simple crystal structure (see insert in Fig.1a) can be considered as model systems representing antiferromagnetic metals that exhibit the quantum critical behavior, the metal-insulator transition and complex charge-spin-orbital-phonon coupling phenomena. With such a variety of intriguing properties, in order to understand the mechanism of the semiconducting and paramagnetic ground state formation and the nature of the metal-insulator transition, comprehensive investigations of the $Tm_{1-x}Yb_xB_{12}$ solid solutions are needed, that will also help to contribute to the understanding of other strongly correlated electron systems. It should also be noted that understanding the mechanisms that determine physical properties of Yb-dodecaborides will expand opportunities for their practical use since the possibility to "tune" the properties of $RB_{12}$ compounds (by sorting through various metals R), in combination with their rigid boron framework that determines hardness, high melting points, mechanical, chemical and thermal stability, make these systems promising for various technical applications. In view of aforesaid, the goal of the present study was to explore the nature of the metal-insulator transition and of the excitations, including the collective modes discovered earlier in $Tm_{0.19}Yb_{0.81}B_{12}$ single crystals [20], by performing the first systematic measurements of the low-temperature AC (infrared) conductivity spectra of this model compound.

**Experimental details.**

For the studies, high quality $Tm_{0.19}Yb_{0.81}B_{12}$ single crystal was grown by inductive zone melting technique, as described in [19], and carefully characterized by the x-ray diffraction, chemical microanalysis and charge transport measurements. For optical measurements, round sample ≈5 mm in diameter was used with plane (within ±1 μm) and polished surface that was etched in the boiling $HNO_3+H_2O$ solution to remove the surface layer with possible structural distortions. At frequencies $\nu$=40-8000 cm$^{-1}$, infrared (IR) reflectivity spectra $R(\nu)$ were measured using Vertex 80V Fourier-transform spectrometer with the gold films deposited on a glass substrate used as reference mirrors. With the J.A. Woollam V-VASE ellipsometer, optical parameters of the sample (real and imaginary parts of the dielectric permittivity and AC conductivity) were directly determined in the frequency interval 3700 – 35000 cm$^{-1}$ with resolution 50 cm$^{-1}$. Measurements with the radiation spot diameter of 2 mm were provided with angles of incidence 65, 70 and 75 degrees. Low-temperature measurements down to $T$=10 K were performed using commercial helium-flow cryostats. From the ellipsometry data, reflection coefficients were calculated using standard Fresnel expressions and merged with the IR reflectivity spectra. The data from [21] were used to extend the reflectivity spectra up to $\approx 400000$ cm$^{-1}$. The obtained broad-band reflectivity was processed as described below. DC conductivity and Hall resistivity of the same crystal were measured using five-terminal scheme at temperatures $T$=2-300 K.

### Experimental results and data analysis.

Figure 1a presents broad-band reflectivity spectra of $Tm_{0.19}Yb_{0.81}B_{12}$ single crystal (dots) at several selected temperatures. The spectra look typically metallic (see, e.g., [22,23]), especially at elevated temperatures, where they reveal pronounced plasma edge (plasma minimum) at $\nu_{pl} \approx 14000$ cm$^{-1}$ and relatively high reflectivity values at lower frequencies. During cooling down, the plasma edge stays practically unchanged, while below $\nu_{pl}$ broad maxima develop in the spectra. We have processed the broad-band reflectivity spectra together with the directly measured with the ellipsometer (at 3700 – 35000 cm$^{-1}$) spectra of real and imaginary parts of dielectric permittivity, introducing the Drude conductivity model to describe the response of free carriers and a set of Lorentzians to model the broad and narrow spectral features. The free carrier response was represented by the Drude expression for the complex conductivity $\sigma^*_D(\nu)$ given as [22,23]

$$\sigma_D^*(\nu) = \frac{\sigma_{DC}}{1 - i\nu/\gamma_D} \qquad (1),$$

where $\sigma_{DC}$ is the DC conductivity and $\gamma_D$ is the charge-carrier scattering rate. In addition to the Drude term, we used the *minimal* (twelve) set of Lorentzians to analyze the collective modes (absorption resonances) and phonons:

$$\sigma^*(\nu) = \frac{0.5 f \nu}{\nu\gamma + i(\nu_0^2 - \nu^2)} \qquad (2).$$

In Eq. (2), $\nu_0$ is the resonance frequency, $f = \Delta\varepsilon\nu_0^2$ is the oscillator strength, $\Delta\varepsilon$ is the dielectric contribution (strength) and $\gamma$ is the damping constant. The examples of least-square fitting results are shown by lines in Fig.1a. The obtained spectrum of the room-temperature IR conductivity $\sigma(\nu)$ is presented in Fig.1b, together with the Drude and Lorentzian components that are shown separately. It is seen that the Drude peak is unusually narrow and corresponds to an unreasonably low scattering rate of carriers for which only estimate can be obtained from the reflectivity spectra fittings, $\gamma_D(290\ K) \leq 10$ cm$^{-1}$. The unusually low value of $\gamma_D$ is a consequence of highly conducting dynamic charge stripes that are present in the $Tm_{0.19}Yb_{0.81}B_{12}$ crystal as was found in [20]. The stripes lead to an "electronic inhomogeneity" of the system: being "immersed" in lesser conducting surrounding matrix, they determine the value of the DC conductivity of the crystal while relatively smaller AC (infrared) conductivity is provided by the analysis of the IR reflectivity of the bulk crystal composed by a "mixture" of stripes and matrix. Confirmation for such interpretation is also found in the Hall effect study of $Tm_{1-x}Yb_xB_{12}$ [19]. It was shown in [19], that additionally to ordinary odd component a transverse even Hall effect was also observed and attributed to the response of the filamentary structure of conductive channels in the dodecaboride matrix.

Fig.2 shows how the spectra of optical conductivity $\sigma(\nu)$ of $Tm_{0.19}Yb_{0.81}B_{12}$ change during cooling down. Strong suppression is seen in the spectra below 1500 cm$^{-1}$ that is mainly due to the weakening of the strength of nine absorption peaks labeled as L1-L3 and L7-L12 in Fig.1b. To characterize quantitatively the temperature behavior and discuss the origin of all L1-L12 peaks, we plot in Fig.3 the temperature dependences of their frequencies, dielectric contributions and damping constants. The broad band centered at $\approx 1800$ cm$^{-1}$ changes only slightly (as can be seen also in Fig.2) when the $Tm_{0.19}Yb_{0.81}B_{12}$ crystal is cooled from 300 K down to 10 K. The band was modeled by a sum of three Lorentzian terms, L4+L5+L6 (see Fig.1b). We associate its origin with the vibrationally coupled localized states near the bottom of the conduction band. These peaks are well-known from previous optical studies of YbB$_{12}$ where the broad bump was described by one Lorentzian with the energy in the range 0.2-0.25 eV depending on the temperature [21,24,25]. In [21,24,25] the authors have also discovered a shoulder in the $\sigma(\nu)$ spectrum located near 40 meV, just on the verge of the energy gap in the density of states of YbB$_{12}$. Similar excitation can be visualized in the spectra of Tm$_{0.19}$Yb$_{0.81}$B$_{12}$ (see Fig.1b) where the corresponding band is represented by the L2+L3 peak. The peaks L9 and L10 with relatively small values of dielectric contributions $\Delta\varepsilon$ should be associated with the quasi-local vibrations of the heavy Yb and Tm ions [20]. The other two low-intensive peaks at ~200 cm$^{-1}$ (L11) and ~400 cm$^{-1}$ (L12) correspond to boron optical phonons which were detected previously in inelastic neutron and Raman spectra (see, e.g., [26,27]).

There are several absorption peaks in the spectra of Tm$_{0.19}$Yb$_{0.81}$B$_{12}$ crystal, denoted as L1, L7 and L8 (see Fig.1b), that were not detected previously in Yb–based dodecaborides. We suggest that these are collective excitations that involve both, the lattice vibrations and the conduction band charge carrier dynamics, as is discussed in the analysis presented below.

The dramatic suppression during cooling down of the AC conductivity of Tm$_{0.19}$Yb$_{0.81}$B$_{12}$ in the range $\nu$=10-1000 cm$^{-1}$ (Fig.2), that is accompanied by marked increase of the DC resistivity and Hall resistance [19], is caused by freezing out of the Drude component and by lowering of the intensities of absorption resonances L1, L2, L3, L7, L8. From Fig. 3a it is seen that during cooling down in the range $T$<100 K, the frequencies of the phonon peaks L9-L12 do not change, while the positions of the collective excitations L4-L6 shift upwards. Simultaneously, a smooth decrease of dielectric contributions of L4-L6 peaks is observed in the entire temperature range, and the $\Delta\varepsilon(T)$ dependences of the L9-L12 peaks demonstrate a series of anomalies at $T$≈150 K (L12) and below 100 K (Fig. 3b). Moreover, below 100 K, a non-monotonic behavior of the damping constants is observed for the peaks L12 and L9-L11 (Fig.3c). Note also a noticeable decrease below 25 K of the damping constant of L9-L11 resonances. According to Fig.3d, the collective modes L2 and L8 become somewhat more rigid at $T$<25 K, while the mode L1 slightly softens. The dielectric contributions of these peaks also demonstrate a rapid decrease below 100 K (Fig. 3e); here, the damping constants of the neighboring L2 and L3 peaks change in opposite directions. At $T$<25 K, the damping constants of the peaks L1, L3, L7 and L8 significantly decrease (Fig.3f). Note that all these resonances have unusually large dielectric contributions (Fig.3e).

**Discussion.**

Before discussing the obtained results, we note firstly that the characteristic temperatures, that determine the change of the charge transport regimes and reconstruction of the charge and spin excitation spectra, were studied in details both, in YbB$_{12}$ and in non-magnetic reference compound LuB$_{12}$. In particular, it was noted in [28] that *fcc* lattice instability caused by the cooperative dynamic Jahn-Teller (JT) effect [29,30] enhances as temperature decreases and leads to a sharp increase in the vibrational density of states (DOS). At the temperatures $T$~$T_E$~150 K, the mean free path of phonons reaches the Ioffe–Regel limit and becomes comparable with their wavelength [31]. Near $T_E$, in addition to the maximum emerging in the vibrational DOS [28], a sharp maximum of the relaxation rate is observed in $\mu$SR experiments in RB$_{12}$ dodecaborides (R = Er, Yb, Lu) and Lu$_{1-x}$Yb$_x$B$_{12}$ solid solutions, as evidenced in [32,33]; it was suggested in [32,33] that the large-amplitude dynamic features arise from atomic motions within the B$_{12}$ clusters.

When discussing the nature of the anomalies observed in LuB$_{12}$ at $T^*$~60 K, the authors of [28,34] concluded that these should be attributed to the order-disorder phase transition, and that at temperatures $T$<$T^*$ the R-ions get frozen in different minima of the double-well potential (see insert in Fig.1a) that are induced by the random distribution of boron vacancies, $^{10}$B and $^{11}$B isotopes and impurities. In this scenario, the barrier height in the double-well potential (see also insert in Fig.1a) is close to the energy corresponding to the cage-glass transition temperature $T^* = 54–65$ K in Lu$^N$B$_{12}$ crystals [28,35] and the disordered state below $T^*$ is represented by a mixture of two components, the crystal (rigid boron covalent cage) and glass (clusters of rare earth ions displaced from their central positions in the B$_{24}$ cubooctahedrons). In [36,37], it was suggested that a defect-induced soft mode is responsible for negative thermal expansion and for the Schottky anomalies of heat capacity at temperatures below 25 K, both having the same origin - formation of two-level tunneling systems based on the metal ions and defects [38]. Besides, according to [37], the existence at intermediate temperatures 60-130 K of negative Gruneisen parameter should be attributed to the presence in the phonon spectra of LuB$_{12}$ of a flat transverse acoustic mode in the superior part of the Brillouin zone (the Einstein temperature $\theta_E \approx 164$ K, k$_B\theta_E \approx 14.1$ meV [39]; here k$_B$ is the Boltzmann constant). In [40], two characteristic temperatures were found for YbB$_{12}$, namely, the temperatures that correspond (i) to the onset of the energy gap formation and the start of the Yb 4f$_{7/2}$ peak shift at $T_E \sim 150$ K, and (ii) to the appearance of the 15 meV structure in photoemission spectra at $T^* \approx 60$ K. In addition, the Kondo peak was observed in YbB$_{12}$ at ~36 meV by hard x-ray photoemission spectroscopy at temperatures below 25 K [41]. The minimum of nuclear spin-lattice relaxation rate (1/$T_1$) at the B site of YbB$_{12}$ was detected suggesting that additional magnetic moments are created below 25 K [42]. A strong increase of electron paramagnetic resonance (EPR) signal was observed in YbB$_{12}$ [43], and the temperature dependence of the EPR amplitude was found to be close to exponential increase with a characteristic temperature ≈18 K. The authors concluded that the EPR results can be understood assuming the existence of Yb$^{3+}$ ion pairs coupled by isotropic exchange interaction, which also interact with neighboring pairs [43] connecting these dimers in the nanosize channels.

Spin excitation spectra of YbB$_{12}$ have been studied in detail in [44-46]. A strong increase with the temperature lowering of the amplitude of the spin-exciton mode localized at 14.5 meV and at around $q$= (1/2,1/2,1/2) in the reciprocal space was detected below 25 K [46]. In the same range of temperatures, $T$<25 K, a transition to the coherent regime of the charge transport was found in Tm$_{1-x}$Yb$_x$B$_{12}$ [19] with a maximum of the Hall mobility and the sign change of the Hall coefficient (see also Fig.4a). In [19], the authors argue that the coherent regime of charge transport in Tm$_{1-x}$Yb$_x$B$_{12}$ should be attributed to the percolation through the network of many-body states in the RB$_{12}$ matrix. Thus, at low temperatures and for $x \geq 0.5$ (on the metallic side of the percolation limit, in the coherent regime), the magnetic system of the intra-gap many-body states should be considered as having a filamentary structure composed of small magnetic clusters of rare earth ions connected by the charge stripes. A similar quasi-two-dimensional character of the spin fluctuation spectrum was revealed in [45] for YbB$_{12}$. Indeed, the peak in the magnetic excitation spectrum of YbB$_{12}$ at ≈14.5 meV was attributed in [45] to antiferromagnetic correlations at the wave vector $q$= (1/2, 1/2, 1/2), and the correlation lengths were estimated both perpendicular and parallel to this direction providing the values $\xi_\parallel = (5.4 \pm 1.4)$ Å and $\xi_\perp = (3.4 \pm 1.1)$ Å and confirming the two-dimensional anisotropy. Moreover, the studies of diffuse neutron scattering in the antiferromagnetic HoB$_{12}$ above the Neel temperature $T_N \approx 7.4$K allowed detection of strong correlations between magnetic moments of Ho$^{3+}$ ions in the paramagnetic state [47,48] that can be explained by appearance of correlated one-dimensional spin chains. These short chains of Ho$^{3+}$ ionic moments located at face diagonals of the elementary unit cell are similar to those in low-dimensional magnets [49] and they can be observed at temperatures up to at least $3T_N \approx 25$ K.

To summarize the above, four characteristic energy scales have been highlighted previously in the studies of RB$_{12}$ compounds that determine the properties of the rare earth dodecaborides at intermediate and low temperatures. (*i*). The JT splitting of the triply degenerate ground electronic state of the B$_{12}$ molecules is estimated to be ~100-200 meV [29]. This means that the cooperative JT dynamics is expected to be observed in the range of frequencies below 1500 cm$^{-1}$. (*ii*). The energy of rattling mode ($\theta_E \sim 110$ cm$^{-1} \sim 14.5$ meV) is very close to the characteristic temperature $T_E \sim 150$ K of the *fcc* lattice instability which develops in RB$_{12}$ crystals leading to strong increase of vibrational DOS when approaching the Ioffe-Regel regime of the lattice dynamics. (*iii*) The order-disorder cage-glass transition at $T^*\approx 60$ K is regulated by the barrier height in the double-well potential for R-ions. Below this temperature the static displacements of $R^{3+}$ ions in the *fcc* lattice are accompanied by emergence of dynamic charge stripes and by charge and spin gaps formation in the matrix of $Tm_{1-x}Yb_xB_{12}$ crystal. (*iv*). At temperatures close to $T_c=25$ K ($k_BT_c\approx 2.2$ meV), a crossover to the coherent regime of the charge transport takes place. It is accompanied by setting up of the low dimensional (1D or 2D) character of the spin and charge fluctuation spectrum and formation of a nanoscale filamentary structure of conduction channels in the matrix of $RB_{12}$.

With the account taken of the above described regimes and the energy scales in the rare earth dodecaborides, we now turn to the discussion of the changes in the microscopic parameters of the $Tm_{0.19}Yb_{0.81}B_{12}$ compound basing on the infrared conductivity measurements. Fig.4a shows the temperature dependences of the Hall coefficient and charge carriers mobility. Vertical dotted lines indicate the temperatures $T_E$, $T^*$ and $T_c$ that are located on the borders of the regimes discussed above. In Fig.4b we present the Arrhenius plot of the oscillator strengths $f=\Delta\varepsilon v_0^2$ of the peaks L2+L3, L7, L8, L1. The activation type dependencies for these quantities are compared with the behavior of the Hall concentration $n_{Hall}$ which demonstrates two linear intervals in the Arrhenius plot with $E_g/2k_B\approx 100$ K ($T\geq 60$ K) and $E_a/k_B\approx 70$ K below $T^*$ (Fig.4b). It is established here that the activation behavior at $T>T^*$ of the oscillator strengths of collective modes L1, L2+L3, L7, L8 is characterized by the energy $E_g/2$. According to [19], the metal–insulator transition with the gap $E_g/2k_B\approx 100$ K, which develops both in $YbB_{12}$ as the temperature decreases and in $Tm_{1-x}Yb_xB_{12}$ as the Yb concentration increases in the range $0< x\leq 1$, is induced by a formation of vibrationally coupled $Yb^{3+}$ pairs randomly distributed in the $RB_{12}$ matrix. Then, similar activation behavior of $n_{Hall}$ and oscillator strengths of the peaks L1, L7, L8, L2, L3 should be considered it terms of a gradual localization of electrons in vicinity of these Yb-dimers (see Fig.4b).

We thus propose that all intra-gap excitations observed in the IR conductivity (with exception of phonons L9-L12, see Fig.1b) are participating in the charge transport at intermediate and low temperatures, and that their total oscillator strength should in the first approximation correspond to the concentration obtained from the Hall effect experiments. Herein, the concentration $n_{Hall}$ of the charge carriers obtained from the Hall effect measurements can be used to estimate the carriers effective mass $m^*$ from the relation $n_{Hall}=\pi m^* f_{tot} e^{-2}$, where $f_{tot}$ is the total oscillator strength of the L1-L3 and L7-L8 peaks and $e$ is electron charge. The so-obtained dependence $m^*(T)$ is presented in Fig.4a. At temperatures $T>T^*$, the value $m^* \sim (5-7)m_0$ corresponds to heavy fermions in the highly correlated electron system $Tm_{0.19}Yb_{0.81}B_{12}$ ($m_0$ is free electron mass). When the temperature is lowered, the effective mass $m^*$ in the disordered phase steeply decreases, and in the coherent regime at $T<T_c\sim 25$ K it acquires the values that are close to $m_0$, that characterize non-equilibrium charge carriers suffering very short-time scattering events. With the relation $\mu(T)=e\tau(T)/m^*(T)$ we can estimate the *mean* scattering time $\tau_\Sigma(T)$ of the charge carriers and compare it with the relaxation time in different scattering channels, L1, L2+L3, L7, L8, and with the inverse width of the quasi-local L12 phonon mode (see Fig.5). The obtained dependences allow us to derive conclusions on the character of charge carriers scattering in different temperature intervals. Above $T_E$, the charge transport in $Tm_{0.19}Yb_{0.81}B_{12}$ is determined by scattering on the collective L1, L3 and L8 modes. After crossing the $T_E$ verge, scattering on the low-frequency L1 mode starts to dominate. Below the transition to the coherent regime at $T_c\approx 25$ K a steep increase of the oscillation frequency of conduction electrons occurs and the role of the L3 mode becomes decisive, providing both a threefold increase of the scattering frequency and decrease of the effective mass to the free electron values $\approx m_0$ (see Figs.4a,5).

Although all above estimates were based on the use of the sum rule, and although, in addition, in the studied unique strongly correlated electronic system the Drude contribution to the free carrier conductivity is very small, we understand that special theoretical efforts are needed to interpret the mechanism that leads to the situation, when collective modes (electron-vibration complexes) determining the contributions to optical conductivity at finite frequencies are also the main components in the low-temperature charge transport of the non-equilibrium charge carriers in $Tm_{0.19}Yb_{0.81}B_{12}$.

**Conclusions**

Our first systematic studies of infrared conductivity of the strongly correlated electron system $Tm_{0.19}Yb_{0.81}B_{12}$ indicate presence of a rich variety of absorption resonances that are not completely screened by free electrons. Tracing the temperature dependences of parameters of the resonances and considering characteristic temperature (energy) scales $T_E$~150 K, $T^*$~60 K and $T_c$~25 K that determine properties of the rare earth dodecaborides at intermediate and low temperatures, we associate the origin of the observed excitations with single-particle dynamics (lattice phonons, Tm and Yb rattling modes) and mixed-type (electron-vibration) collective modes which are responsible both for the gap opening ($E_g/2k_B \approx 100$ K) and the emergence of the manybody intra-gap states ($E_a/k_B \approx 70$ K) in matrix of the dodecaborides. We argue that the Jahn-Teller instability of the boron sub-lattice with the characteristic frequency ~1500 cm$^{-1}$ is among the main factors that determine the complexity of physical properties of these dodecaborides. Temperature-dependent effective mass of free charge carriers is estimated and a picture of multiple relaxation channels at different temperatures is presented. The obtained results will be helpful in understanding the mechanism of the ground state formation and the nature of the metal-insulator transition in $Tm_{1-x}Yb_xB_{12}$, as well as in other strongly correlated electron systems including high temperature superconductors, manganites, etc.

**Acknowledgements**

The research was supported by the RSF grant №17-12-01426 and by the Ministry of Science and Higher Education of the Russian Federation (Program 5 top100). Authors acknowledge the Shared Facility Center at P.N. Lebedev Physical Institute of RAS for using the equipment and the state assignment of the Ministry of Science and Higher Education (theme #0023-2019-0005).

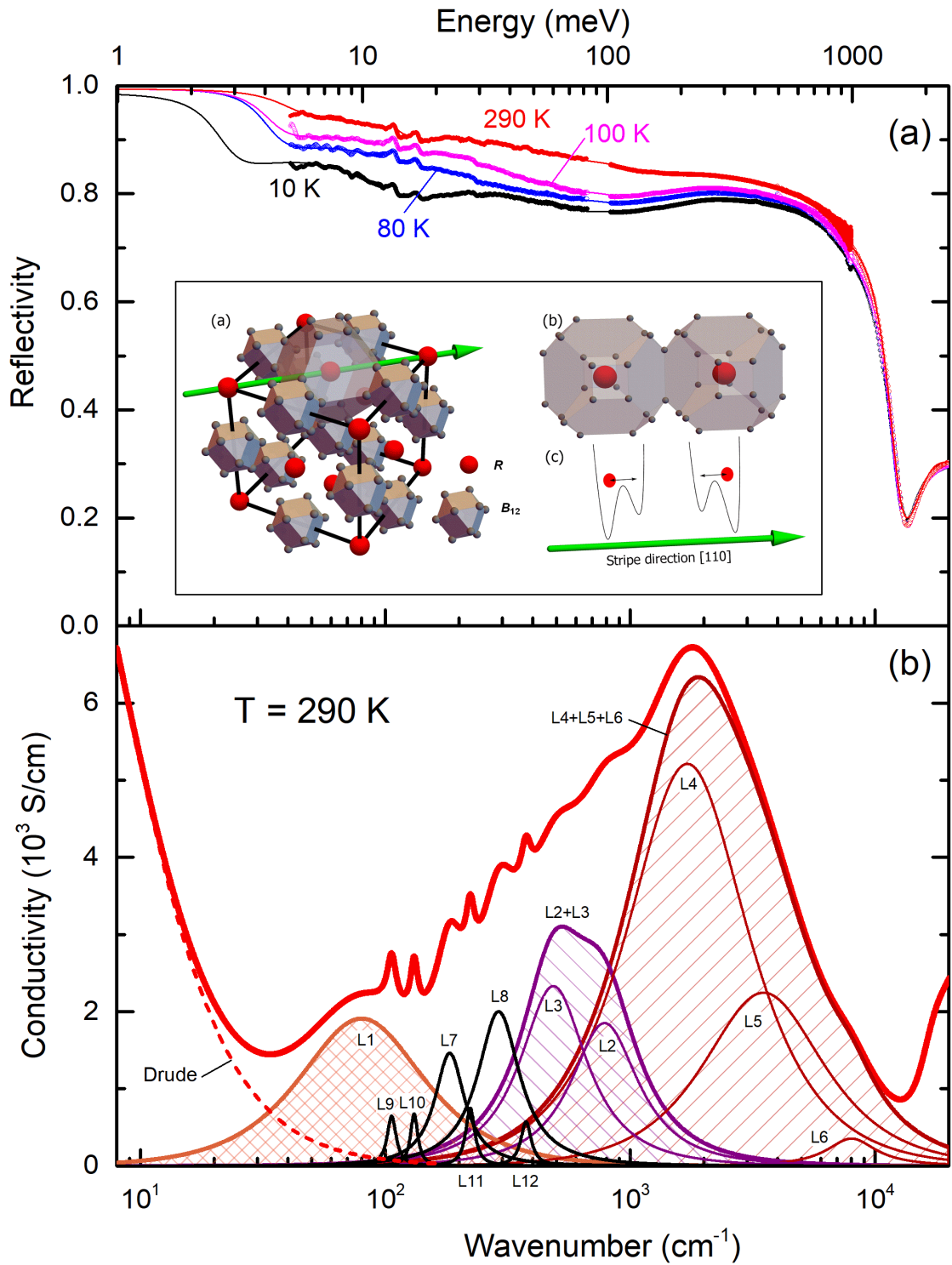

Fig.1. (a) Room temperature spectra of reflection coefficient of $Tm_{0.19}Yb_{0.81}B_{12}$ single crystal (dots). Solid lines show the results of least-square fitting of the spectra with the Drude term Eq.(1) and Lorentzian terms Eq.(2) as described in the text. The obtained spectrum of real part of conductivity is presented in panel (b), where the Drude (dashed line) and the Lorentzian contributions are shown separately. Inset in panel (a) presents the crystal structure of $RB_{12}$ (left) with the fragment (right-up) consisting of the couple of truncated boron cuboctahedrons centered by rare earth ions. Lower right part of the inset demonstrates schematic view of vibrations of R-ions in the double-well potential. Green arrow shows the direction of dynamic charge stripes in the lattice of dodecaboride.

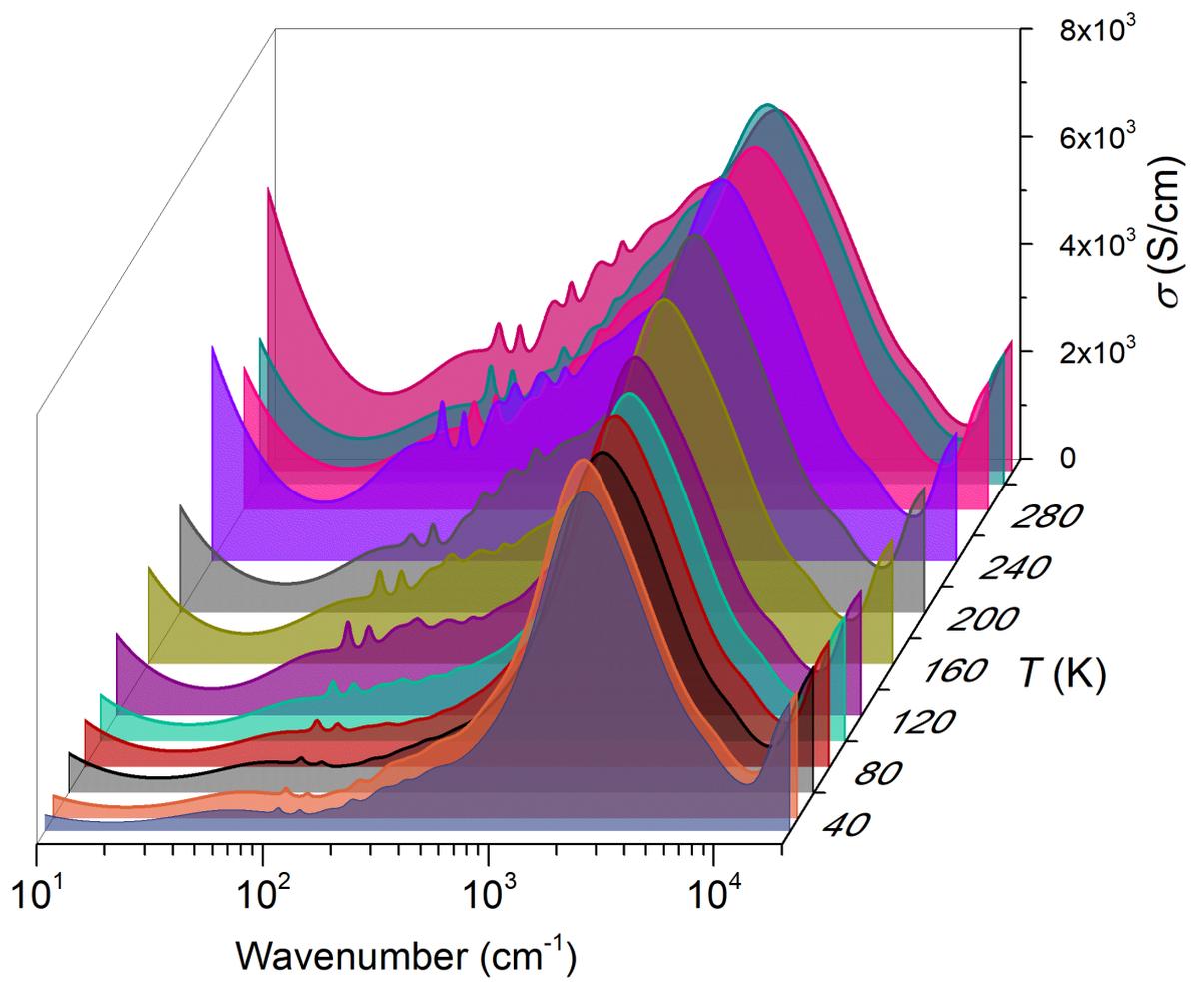

Fig.2. Evolution during cooling down of the spectrum of infrared conductivity of $Tm_{0.19}Yb_{0.81}B_{12}$ single crystal.

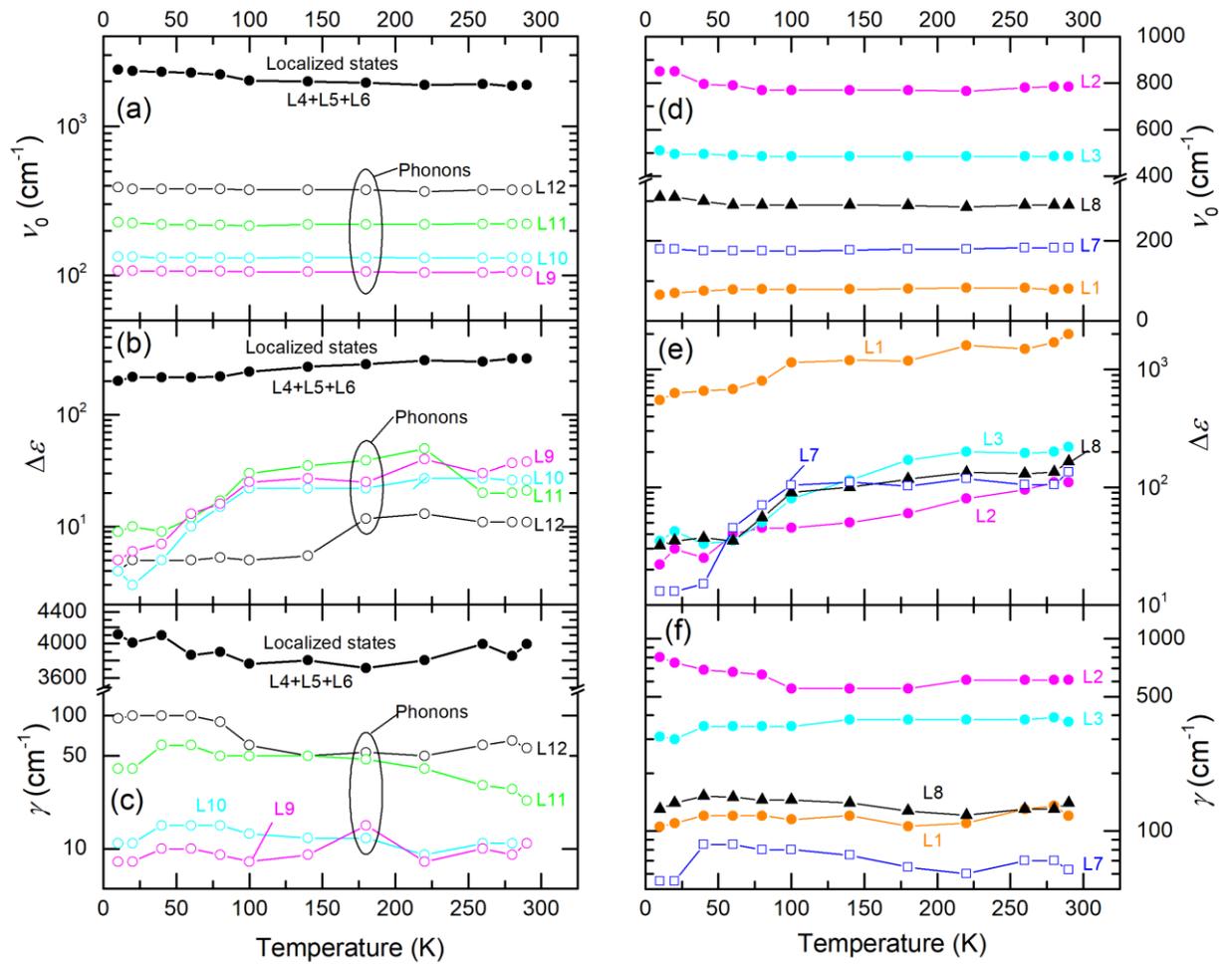

Fig.3. Temperature dependences of parameters of absorption peaks observed in the spectra of infrared conductivity of $Tm_{0.19}Yb_{0.81}B_{12}$ single crystal and modeled using Lorentzian terms, Eq.(2): frequencies (a), (d), dielectric contributions (b), (e), damping constants (c), (f). The peaks are labeled according to Fig.1b. Peaks associated with phonons and localized states are specified.

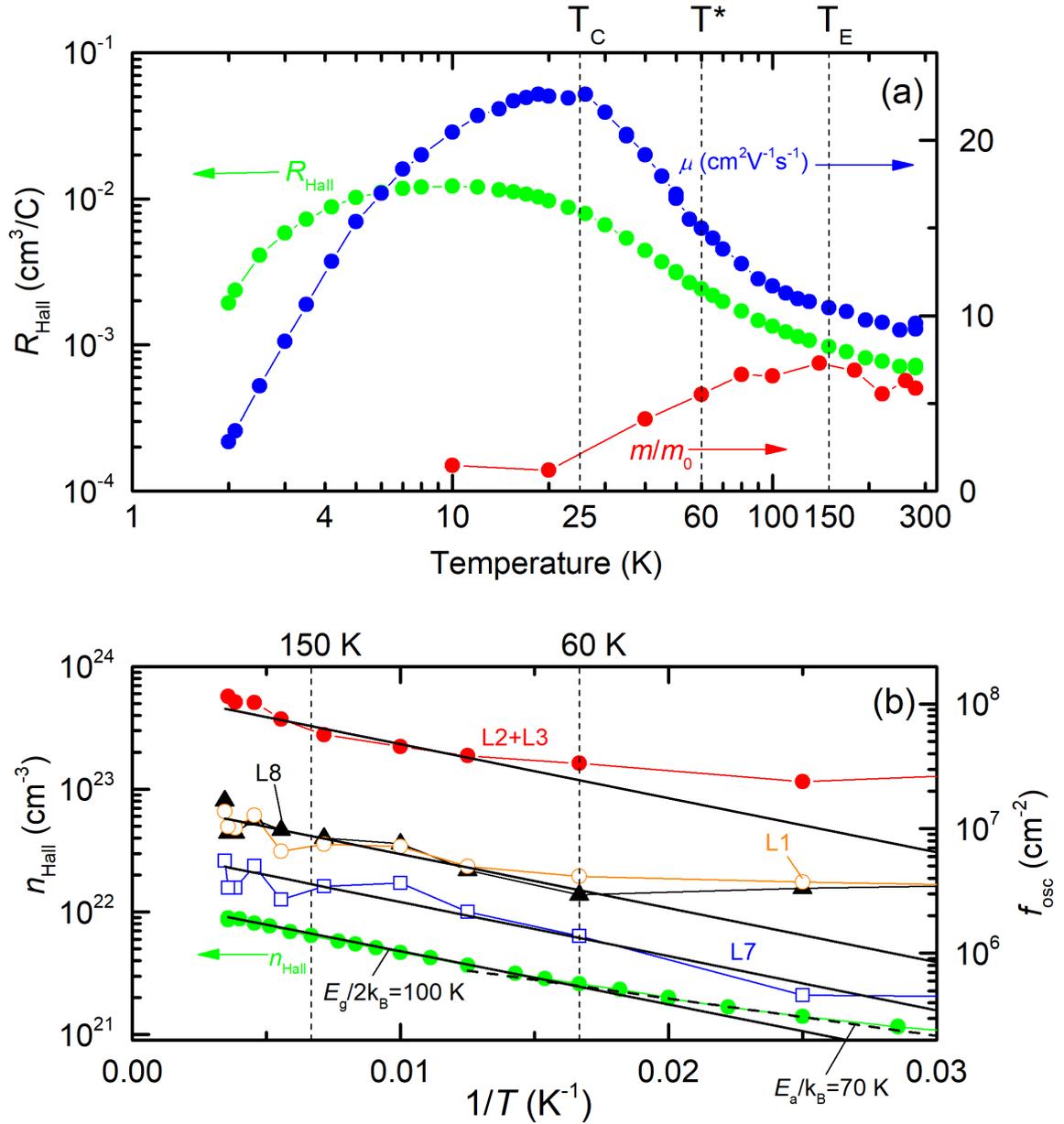

Fig.4. (a) Temperature dependences of the Hall constant $R_{Hall}$, mobility $\mu$ and effective mass $m^*/m_0$ of charge carriers in $Tm_{0.19}Yb_{0.81}B_{12}$ single crystal. Effective mass is calculated as described in the text. (b) Arrhenius plot of the temperature dependences of concentration of charge carriers in $Tm_{0.19}Yb_{0.81}B_{12}$ single crystal ($n_{Hall}$) calculated from the Hall constant, and of oscillator strengths of peaks modeled with the Lorentzian terms, Eq.(2): L1, L2 and L3 (summed), L7 and L8. The peaks are labeled according to Fig.1b. Solid and dashed lines in panel (b) correspond to activated behaviors $\sim\exp[-E_g/(2k_B)]$ with activation temperatures $E_g/(2k_B)=100$ K and 70 K, respectively (as indicated). Vertical dotted lines at the characteristic temperatures $T_c$, $T^*$ and $T_E$ are located at the transitions to coherent regime of charge transport ($T_c\sim25$ K), to the disordered state of the rare earth ions (cage-glass transition at $T^*\sim60$ K) and on the boundary of the *fcc* lattice instability (Ioffe-Regel regime, $T_E\sim150$ K).

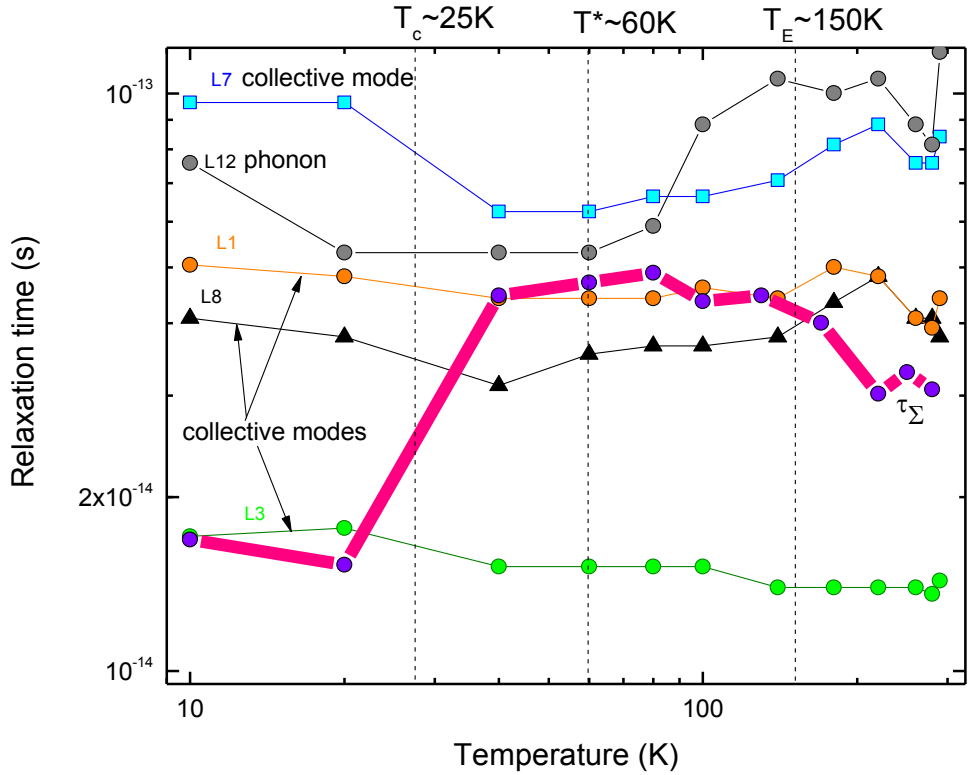

Fig.5. Temperature dependences of relaxation times of absorption peaks observed in the infrared spectra of $Tm_{0.19}Yb_{0.81}B_{12}$ single crystal and labeled as L1, L3, L7, L8 and L12 (see Fig.1b). Thick line corresponds to relaxation time $\tau_\Sigma$ of charge carriers calculated from their mobility and effective mass shown in Fig.4a. Peaks associated with phonons and collective states are specified. Vertical dotted lines at the characteristic temperatures $T_c$, $T^*$ and $T_E$ are located at the transitions to coherent regime of charge transport ($T_c \sim 25$ K), to the disordered state of the rare earth ions (cage-glass transition at $T^* \sim 60$ K) and on the boundary of the *fcc* lattice instability (Ioffe-Regel regime, $T_E \sim 150$ K).